\documentclass[12pt,a4paper]{article}

\newcommand{\la}{\lambda}
\newcommand{\al}{\alpha}

\begin{document}

\begin{flushright}
hep-th/0103145
\end{flushright}
\vspace{1.8cm}

\begin{center}
\textbf{\Large Dielectric D0-Branes in Near-Horizon D-Brane \\
  Backgrounds and D-Strings under Electric RR Fluxes}
\end{center}
\vspace{1.6cm}
\begin{center}
Shijong Ryang
\end{center}

\begin{center}
\textit{Department of Physics \\ Kyoto Prefectural University of
 Medicine \\ Taishogun, Kyoto 603-8334 Japan} \\
\texttt{ryang@koto.kpu-m.ac.jp}
\end{center}
\vspace{2.8cm}
\begin{abstract}
Using the nonabelian Dirac-Born-Infeld action with the Wess-Zumino term
that is constructed in consistent with T duality we examine the Myers
dielectric effect for multiple D0-branes in the near-horizon geometry of
D4-branes. The effect in the curved spacetime is also confirmed  by the
dual formulation based on the abelian Dirac-Born-Infeld action of
a D2-brane. Putting a system of muliple D-strings in 
the external electric RR five-form flux, we construct a noncommutative
non-BPS solution where the D-strings expand into a spherical D3-brane.
We discuss the external field dependence of the funnel-like 
or wormhole solution.
\end{abstract}
\vspace{3cm}
\begin{flushleft}
March, 2001
\end{flushleft}
\newpage
\section{Introduction}

Recently there have been interesting observations in string theory that
the objects that are naively point-like become the expanded branes 
under certain background fluxes. A system of the multiple D0-branes in
the electric RR four-form field strength can develop a dipole moment
under the corresponding three-form potential \cite{RCM,TR}. Its lowest 
energy configuration is described by a noncommutative two-sphere that is
regarded as a single spherical D2-brane.  This Myers dielectric effect 
has been demonstrated in the flat spacetime by analyzing 
the nonabelian Dirac-Born-Infeld (DBI) action with the Wess-Zumino 
term that is constructed by means of T duality.  
On the other hand the behaviors of a graviton propagating in the
spherical part of product space $AdS_m \times S^n$ 
have been studied \cite{MST}.
The graviton that is point-like at low angular momentum blows up into the
expanded brane state for large angular momentum. The size of giant 
graviton increases with increasing angular momentum till it becomes
the maximal size specified by the radius of the surrounding sphere, 
which provides the maximal bound of angular momentum. 

The magnetic analogue of the Myers dielectric effect has been studied and
related with the giant graviton \cite{DTV}.
When a D$p$-brane in the near-horizon geometry produced by the multiple
D$(6-p)$-branes is placed in a magnetic RR $(p+2)$-form field strength,
it expands into a D$(p+2)$-brane which behaves like a massless particle
with a bounded angular momentum. There have been several studies about the
Myers dielectric effect \cite{PS,TV,GJS,IB}.
In the type IIB string theory the polarization
of D3-branes placed in the electric seven-form field strength has been
investigated from the view point of dual description where the 
abelian DBI action of a five-brane in the $AdS_5 \times S^5$ curved 
spacetime has been analyzed \cite{PS}.
The extensions to other noncommutative fuzzy manifolds such as 
$S^2 \times S^2, CP^2$ have been performed \cite{TV}.
For a system of multiple D-instantons in the presence of electric RR
five-form field strength in the flat spacetime a stable noncommutative
solution has been found \cite{GJS}.

Based on the nonabelian DBI theory in consistent with T duality 
the noncommutative solutions have been constructed in a system of $N$
coincident D-strings in the flat spacetime as well as the D3-brane
background \cite{CMT}. The BPS solutions describe 
the D-strings expanding out in
a funnel-like geometry to become a spherical D3-brane. 
This expanding behavior is distinct from the Myers dielectric effect
since it is observed although the external electric RR flux is absent.
It occurs owing to the nonabelian nature of the world-volume theory.

For the well understood abelian DBI theory of a D2-brane 
with a world-volume electric flux that is placed 
in the background of the electric RR four-form field
strength, the interesting static solultions have been found where
the fundamental strings can be viewed as D2-branes whose world-volumes
have collapsed to string-like configurations \cite{RE}.
Further for this system the other type of finite-energy solution has
been presented \cite{PTM}.

Working on the nonabelian DBI theory proposed in Ref. \cite{RCM}
we will construct a noncommutative solution for a system of multiple
D0-branes in the near-horizon geometry of D4-branes,
which is placed in the constant electric RR four-form flux.
In the dual description based on the abelian DBI theory of a 
single D2-brane in the same D4-brane background under the electric
RR flux the corresponding solution will be analyzed. Through the two 
approaches we will examine the dielectric effect in the curved spacetime.
We will study the same type of nonabelian 
DBI theory for a system of multiple D-strings in the flat 
spacetime. When the constant electric RR five-form
flux is turned on we will demonstrated how a noncommutative solution
appears. This solution will be compared with the BPS funnel-like 
solution in the abcence of external electric RR flux. 

\section{Dielectric effects in the near-horizon geometries}

The nonabelian DBI action describing $N$ coincident D$p$-branes 
in consistent with T duality is given by
\begin{equation}
S_{DBI} = - T_p \int d^{p+1}\sigma \mathrm{STr} (e^{-\varphi}
\sqrt{-\det (P[ E_{ab} + E_{ai}(Q^{-1} - \delta)^{ij}E_{jb}] +
\la F_{ab}) \det (Q^i_{\;j})} )
\label{dbi}\end{equation}
with $E_{\mu\nu} = G_{\mu\nu} + B_{\mu\nu}$ and $Q^i_{\;j} \equiv
\delta^i_{\;j} + i\la [\Phi^i, \Phi^k]E_{kj}$ where the symmetrized trace
prescription is used and $T_p$ is the tension of D$p$-branes \cite{RCM}.
The world-volume scalars $\Phi^i, i = p+1, \cdots, 9$ are $N \times N$
matrices in the adjoint representation of the $U(N)$ world-volume gauge
symmetry. The pull-back of the bulk spacetime tensors to the D$p$-brane
world-volume is expressed in terms of the covariant derivatives of 
the nonabelian scalars as
\begin{equation}
P[E_{ab}] = E_{ab} + \la E_{ai}D_b\Phi^i + \la E_{ib}D_a\Phi^i +
\la^2 E_{ij}D_a\Phi^i D_b\Phi^j,
\end{equation}
where $a =0,1, \cdots, p$ and we have chosen a static gauge and the 
transverse displacements are given by $X^i = \la \Phi^i$ with 
$\la = 2\pi l^2_s$. The Wess-Zumino action is described by
\begin{equation}
S_{WZ} = T_p \int \mathrm{STr}(P[e^{i\la i_{\Phi}i_{\Phi}}(\sum
C^{(n)}e^B)] e^{\la F}),
\label{wz}\end{equation}
where $i_{\Phi}$ denotes the interior product by $\Phi^i$ regarded as a
vector in the transverse space, acting on the $n$-form RR potential 
$C^{(n)}$. The interaction terms in the Wess-Zumino action
show that the $N$ D$p$-branes couple
to RR potentials of degree $p+3$ and above, besides having the well
understood couplings to RR potentials of degree equal to and
smaller than dimensions of the world-volume.

We consider the $N$ coincident D0-branes sitting in the background
of the near-horizon geometry of $N_4$ D4-branes in the type IIA string 
theory. Turning on a constant electric RR four-form field strength 
associated with D2-brane charge we  examine the dielectric 
effect for the D0-branes. The metric and dilaton for this background
are given by
\begin{eqnarray}
ds^2 &=& H^{-1/2}(-dt^2 + \sum_{n=1}^{4}(dX^n)^2 ) + H^{1/2}
\sum_{i=5}^{9}(dX^i)^2,
\nonumber \\ e^{\varphi} &=& H^{-1/4},
\end{eqnarray}
where $H = (L/r)^3$ and $L^3 = \pi g_sl^3_sN_4$. Through the 
parametrization
\begin{eqnarray}
(X^5, X^6) &=& r\sqrt{1-\rho^2} (\cos \phi, \sin \phi), \nonumber \\
(X^7, X^8, X^9) &=& r\rho(\cos\theta, \sin\theta\sin\psi, 
\sin\theta\cos\psi)
\label{cor}\end{eqnarray}
the metric reduces to
\begin{eqnarray}
ds^2 &=& H^{-1/2}( -dt^2 + (dX^n)^2 ) + H^{1/2}( dr^2 + \frac{r^2}
{1- \rho^2}d\rho^2 + r^2(1-\rho^2)d\phi^2 \nonumber \\
&+& r^2\rho^2d\theta^2 + r^2\rho^2\sin^2\theta d\psi^2 ).
\end{eqnarray}
We take the transverse electric RR four-form field strength to be
\begin{equation}
F^{(4)}_{t,6+i,6+j,6+k} = -2f\epsilon_{ijk},\; \mathrm{for}\; 
i,j,k \in \{1,2,3\}
\label{str}\end{equation}
with a constant $f$ and vanish for otherwise so that the transverse 
scalars $\Phi^{6+i} = X^{6+i}/\la, i=1,2,3$ are relevant.
Though we intend to study the static solution for the D0-branes
we leave $\phi$ arbitrary time-dependent to construct the Hamiltonian
of this system later. We construct the 
static solution where $X^5, X^6$ are 
commutative variables and $X^{6+i}\; ( i=1,2,3 )$ are noncommutative ones
so that $r, \rho, \phi$ are regarded to be proportional to the 
$N \times N$ identity matrix $I_N$. Since $[\Phi^5,\Phi^6]=0$ and
$P[G_{00} + G_{0i}(Q^{-1} - \delta)^{ij}
G_{j0}] = G_{00} + \la^2\partial_0\Phi^iG_{ij}(Q^{-1})^{jk}G_{kl}
\partial_0\Phi^l$, the DBI action (\ref{dbi}) for the D0-branes reads
\begin{equation}
S_{DBI} = - T_0 \int dt \mathrm{STr}(e^{-\varphi}\sqrt{(H^{-1/2} -
H^{1/2}r^2(1-\rho^2)\dot{\phi}^2)(1 - \frac{\la^2H}{2}
[\Phi^i, \Phi^j]^2)} ),
\label{act}\end{equation}
where $\Phi^{6+i}$ have been replaced by $\Phi^i$ for convenience of
notation. The Wess-Zumino action (\ref{wz})
produces an interaction $i(\la^2T_0/3)\int dt\mathrm{Tr}
(\Phi^i\Phi^j\Phi^k)F^{(4)}_{t,6+i,6+j,6+k}$.
The equation of motion for $\Phi^i$ is obtained through
\begin{equation}
\mathrm{STr}\left( \left( H\frac{[\Phi^j,[\Phi^j,\Phi^i]]}
{\sqrt{1 - \frac{\la^2H}{2}[\Phi^i,\Phi^j]^2}} 
+ if\epsilon_{ijk}[\Phi^j,\Phi^k] \right)\delta \Phi^i \right) = 0.
\end{equation}
Substitution of the ansatz 
\begin{equation}
 \Phi^i = R\al^i
\label{ans}\end{equation}
with $\al^i$ SU(2) generators in the irreducible $N \times N$ 
representation yields the solutions
\begin{equation}
R^2_{\pm} = \frac{H^2 \pm \sqrt{H^4 - \la^2CHf^4}}{2\la^2CHf^2}
\label{rso}\end{equation}
with the quadratic Casimir of this representation $C = N^2 - 1$.
Due to the RR external flux the $N$ coincident D0-branes expand into the
form of a noncommutative two-sphere whose size is measured by the physical
radius $\tilde{R}$ which is defined by  $\tilde{R}^2 = \la^2 
\sum_{i=1}^3 \mathrm{Tr}(\Phi^i)^2/N = \la^2CR^2$. Combining the solutions
(\ref{rso}) with a constraint $\sum_{i=1}^3(X^{i})^2 = r^2
\rho^2 \mathrm{I}_N$ we see that $\rho$ is specified as
\begin{equation}
\rho_{\pm} = \frac{\tilde{R}_{\pm}}{r}
\label{rho}\end{equation}
for the fixed $r$. 

Alternatively substituting the ansatz (\ref{ans}) directly into the 
starting DBI action (\ref{act}) with the Wess-Zumino term and replacing
the $\al^i\al^i$ by $C\mathrm{I}_N$ approximately we have the momentum
$P_{\phi}$ conjugate to $\phi$ and construct the Hamiltonian 
of this system
\begin{equation}
H_{D0} = H^{-1/4}\left( \frac{P^2_{\phi}}{H^{1/2}r^2(1-\rho)} +
(T_0 NH^{1/4})^2(1+(2\la R^2)^2HC) \right)^{1/2} - \frac{4}{3}
T_0\la^2 NCR^3f.
\label{ham}\end{equation}
We can confirm that extremizing the Hamiltonian with respect to $R$ for
the static case gives an equation of motion which 
determines the radius of the puffed two-sphere and reproduces the 
solutions (\ref{rso}). For the weak external field $R_-$ becomes
$f/2H$ due to the near-horizon geometry $H\gg 1$. This $R_-$ solution,
when $H$ is replaced by 1, agrees with the solution for the flat 
spacetime in Ref. \cite{RCM}. Whether the $R_+$ solution exists or not
will be argued later from the view point of the dual description.
The position dependent mass term in (\ref{ham}) for the $R_-$ solution
is rewritten as
\begin{equation}
\sqrt{g_{tt}}((NT_0e^{-\varphi})^2 + (T_2e^{-\varphi}S_A)^2 )^{1/2},
\end{equation}
where $\sqrt{g_{tt}}$ is the redshift factor and $S_A = 
H^{1/2}4\pi \tilde{R}^2_-$ is the physical
surface area of the puffed two-sphere. This expression implies
that the $R_-$ solution gives the nonmarginal bound state consisting of
$N$ D0-branes and a spherical D2-brane.

On the other hand there is a dual formulation of the same system. 
It is described by the abelian DBI action of a single D2-brane with $N$ 
units of U(1) magnetic flux that represent $N$ D0-branes bound to the
D2-brane in the near-horizon D4-brane background.
From (\ref{str}) we make a convenient gauge choice to represent the RR
three-form potential as $C_{t,6+i,6+j} = 2f\epsilon_{ijk}X^{6+k}/3$.
Through the static gauge $\sigma^0 = t = X^0, \sigma^1 =\theta, 
\sigma^2 = \psi$ this potential together with (\ref{cor}) yields 
a well understood coupling in the Wess-Zumino action
\begin{equation} 
T_2\int P[C^{(3)}] = T_2 \int dtd\theta d\psi \frac{2f}{3}(r\rho)^3
\sin \theta.
\end{equation}
For this system where a D2-brane wraps a two-sphere of radius $r\rho$ in
the $X^{7,8,9}$ directions and moves to the $\phi$ direction, the 
Hamiltonian is also derived as 
\begin{equation}
H_{D2} = H^{-1/4}\left( \frac{P^2_{\phi}}{H^{1/2}r^2(1-\rho)} + 
(4\pi T_2H^{1/4})^2( (H^{1/2}r^2\rho^2)^2 + \frac{N^2\la^2}{4})
\right)^{1/2} - \frac{8\pi}{3} T_2 (r\rho)^3f.
\end{equation}
The two-sphere whose surface area is given by $\int d\theta 
d\psi \sqrt{\det G_{\Omega_2}}=H^{1/2}4\pi r^2\rho^2$,  
is embedded in the transverse four-sphere in 
the $X^i (i=5,\cdots,9)$ directions. The static equilibrium equation 
with respect to $\rho$ is given by
\begin{equation}
\frac{Hr\rho}{\sqrt{H(r\rho)^4 + \frac{N^2\la^2}{4}}} = f,
\end{equation}
whose solutions are expressed as 
\begin{equation}
\rho^2_{\pm} = \frac{H^2 \pm \sqrt{H^4 - \la^2N^2Hf^4}}{2r^2Hf^2}.
\end{equation}
Up to $1/N^2$ corrections through (\ref{rho}) these solutions for the
dual abelian description of a D2-brane agree with the previous solutions
(\ref{rso}) for the nonabelian description of D0-branes.
But the solutions exist only when $\rho^2_{\pm} < 1$ is satisfied.
For the weak external field the solutions can be approximately given by
$\rho^2_+ \simeq H/r^2f^2, \rho^2_- \simeq \la^2N^2f^2/4r^2H^2$ in
the near-horizon geometry $0<r \ll L$. Since these expressions can be
rewritten as $\rho^2_+ \simeq H^{5/3}/L^2f^2, \rho^2_- \simeq
(\la^2N^2f^2/4L^2)H^{-4/3}$, the $\rho_+$ solution does not appear while
the $\rho_-$ solution exists. 

\section{Dielectric D-strings}

We consider the $N$ coincident D-strings for the type IIB string theory in
the flat background, whose nonabelian DBI action is given by
\begin{equation}
S_{DBI} = - T_1 \int d^2\sigma \mathrm{STr} \sqrt{-\det (\eta_{ab} + 
\la^2\partial_a\Phi^iQ^{-1}_{ij}\partial_b\Phi^j)\det(Q^{ij})},
\label{dst}\end{equation}
where the world-volume gauge field is taken to be zero. Choosing the 
static gauge $\tau = t =X^0, \sigma =X^9$ and expanding this nonabelian
action in $\la$ we have
\begin{equation}
S_{\la} = -T_1 \int dtd\sigma ( N + \frac{\la^2}{2}\mathrm{Tr}
(\partial^a\Phi^i\partial_a\Phi^i + \frac{1}{2}[\Phi^i,\Phi^j]
[\Phi^j,\Phi^i]) + \cdots ).
\label{lin}\end{equation}
We are concerned with the behavior of $N$ D-strings when we turn on a 
constant electric RR five-form field strength $F^{(5)}$ that is associated
with D3-brane charge. In this case the Wess-Zumino action (\ref{wz}) 
yields an interaction term
\begin{equation}
i\frac{\la^2T_1}{3}\int dtd\sigma \mathrm{Tr}(\Phi^i\Phi^j\Phi^k)
F^{(5)}_{t,\sigma,5+i,5+j,5+k}\; ,
\label{fiv}\end{equation}
where $F^{(5)}_{t,\sigma,5+i,5+j,5+k}
 = -2f\epsilon_{ijk}$ only for $i,j,k \in \{1,2,3\}$ and
$\Phi^{5+i}$ have been again replaced by $\Phi^i$. From 
the linearized action (\ref{lin}) and 
(\ref{fiv}) the equation of motion for the nonabelian scalars is given by
\begin{equation}
\partial^a\partial_a \Phi^i = [\Phi^j,[\Phi^j,\Phi^i]] + 
if\epsilon_{ijk}[\Phi^j,\Phi^k].
\end{equation}
For the static solution we use also the ansatz (\ref{ans}) whose $R$ here
has a $\sigma$ dependence to derive an equation
\begin{equation}
\frac{d^2R}{d\sigma^2} = 8R^3 - 4fR^2.
\end{equation}
The integration of it gives
\begin{equation}
\frac{dR}{d\sigma} = \pm \sqrt{4(R^4 - R_0^4) - \frac{8}{3}fR^3},
\label{dif}\end{equation}
where $R_0^4$ is an integration constant. The further integration with an
integration constant $\sigma_{\infty}$ yields the general solution
\begin{equation}
\sigma = \sigma_{\infty} \pm \frac{1}{2}\int_{\infty}^{R} \frac{dR}
{\sqrt{R^4-R_0^4 - \frac{2}{3}fR^3}}.
\label{ges}\end{equation}
If $R_0$ is chosen to be zero, $R$ can be obtained explicitly in
terms of $\sigma$ as
\begin{equation}
R = \frac{1}{\mp2(\sigma - \sigma_{\infty}) - \frac{2}{3}
f(\sigma - \sigma_{\infty})^2}.
\label{ers}\end{equation}
When there is no external background field $f=0$, 
this solution reduces to the
supersymmetric funnel solution where the D-strings open up into an
anti-D3-brane/D3-brane at $\sigma = \sigma_{\infty}$ and the spherical
three-brane becomes more and more string-like as $\sigma \rightarrow
\mp \infty$ \cite{CMT}. Our solution (\ref{ers}) with (\ref{dif})
shows that $dR/d\sigma$ vanishes at $\sigma = \sigma_{\infty} \mp
3/2f$ where $R$ takes $2f/3$. Due to the non-zero electric external field
a three-brane bubble with a noncommutative $R^1 \times S^2$ structure is
produced from a system of $N$ D-strings. The spherical 
anti-D3-brane/D3-brane with the infinite radius at $\sigma = 
\sigma_{\infty}$ shrinks as one moves along the $\mp \sigma$ direction
to become the smaller three-brane. Combining the two minus/plus solutions
in (\ref{ers}) a wormhole solution can be constructed. Through the plus
solution the D3-brane with the infinite radius at $\sigma = 
\sigma_{\infty}$ shrinks as $\sigma$ increases and the radius takes the
minimum value $2f/3$ at $\sigma = \sigma_{\infty} + \Delta\sigma$ with
$\Delta\sigma = 3/2f$. This plus solution is continued to the minus
solution which grows into an anti-D3-brane at $\sigma = \sigma_{\infty}
+2\Delta\sigma$.

Here without using the linearized world-volume action we manipulate
the full DBI action. The direct substitution of our ansatz into 
(\ref{dst}) and (\ref{fiv}) yields
\begin{equation}
S = -T_1 \int dtd\sigma \mathrm{STr} (\sqrt{(1+\la^2\al^i\al^i(R')^2)
(1 + 4\la^2\al^j\al^jR^4)} - \frac{4}{3}\la^2R^3f\al^i\al^i ).
\label{ful}\end{equation}
Making an approximation to replace the $\al^i\al^i$ by $C\mathrm{I}_N$
we get a full equation of motion for $R$
\begin{equation}
\frac{d}{d\sigma} \left( \frac{R'\sqrt{1+4\la^2CR^4}}
{\sqrt{1+\la^2C(R')^2}}\right) - 8R^3 
\sqrt{\frac{1+\la^2C(R')^2}{1+4\la^2CR^4}} + 4R^2f = 0,
\end{equation}
which describes the evolution of radius as one moves along the 
D-string worldsheet.
This equation can be cast into
\begin{equation}
\frac{d}{d\sigma}\left( \sqrt{\frac{1+4\la^2CR^4}{1+\la^2C(R')^2}}
- \frac{4\la^2C}{3}fR^3 \right) = 0.
\label{cas}\end{equation}
Using an integration constant $A$ we have a first integral
\begin{equation}
\la\sqrt{C}\frac{dR}{d\sigma} = \pm \frac{\sqrt{1+4\la^2CR^4-
(\frac{4\la^2C}{3}fR^3 + A)^2}}{\frac{4\la^2C}{3}fR^3 + A}.
\label{fir}\end{equation}
We will analyze this equation in the $A=1$ case which corresponds to the
previous $R_0 = 0$ case, because in this case (\ref{fir}) is compared
with (\ref{dif}) in the small $f$ limit. From (\ref{fir}) we obtain
the solutions implicitly with an integration constant $\sigma_c$
\begin{equation}
\sigma = \sigma_c \pm \la\sqrt{C}\int_{R_c}^{R}dR \frac{R^3 + 
\frac{3}{4\la^2Cf}} {R^{3/2}\sqrt{-R^3 + \al R + \beta}},
\label{ims}\end{equation}
where $\al = 9/4\la^2Cf^2, \beta = -3/2\la^2Cf$ and $R$ takes $R_c$ at
$\sigma = \sigma_c$. When $f^4 < 3/4\la^2C$, the third order equation
$R^3 - \al R - \beta =0$ has the following three real roots
\begin{eqnarray}
R_1 &=&-2\left(\frac{3}{4\la^2Cf^2}\right)^{1/2}\cos\frac{\psi}{3},
\; R_2 =  \left(\frac{3}{4\la^2Cf^2}\right)^{1/2}(\cos\frac{\psi}{3}
-\sqrt{3}\sin\frac{\psi}{3}), \nonumber \\
R_3 &=&  \left(\frac{3}{4\la^2Cf^2}\right)^{1/2}(\cos\frac{\psi}{3}
+ \sqrt{3}\sin\frac{\psi}{3})
\label{rad}\end{eqnarray}
with $\cos\psi = \sqrt{4\la^2C/3}f^2$ for $0< \psi <\pi/2$, which obey
$R_1 < 0 < R_2 < R_3$ and $R_1 + R_2 + R_3 = 0$. 
The three roots can be expanded in a small
parameter $x = 2\la\sqrt{C}f^2/\sqrt{3}<1$ as
\begin{eqnarray}
R_1 &=& - \frac{f}{x}(\sqrt{3} + \frac{1}{3}x - \frac{\sqrt{3}}{18}x^2
+ \frac{4}{81}x^3 + \cdots ), \nonumber \\
R_2 &=& \frac{f}{x} ( \frac{2}{3}x + \frac{8}{81}x^3 + \cdots ), \\
R_3 &=&  \frac{f}{x}(\sqrt{3} - \frac{1}{3}x - \frac{\sqrt{3}}{18}x^2
- \frac{4}{81}x^3 + \cdots ). \nonumber
\end{eqnarray}
In the small $f$ limit $R_2$ approaches to $2f/3$ and $R_3$ grows as
$3/2\la\sqrt{C}f$. Therefore if we choose $R_c = R_3$, the expression 
(\ref{ims}) corresponds to (\ref{ges}). The limiting value $R_2 = 2f/3$
agrees with the previous minimum radius in (\ref{ers}).
In the weak field region $f^4 < 3/4\la^2C$ the radius takes a range
$R_2<R<R_3$ due to $-R^3 + \al R + \beta > 0$. The solution
for the minus sign in (\ref{ims}) with $R_c = R_3, \sigma_c = \sigma_3$
implies that the D-string with a geometry $R^1 \times S^2$
opens up into a D3-brane with a finite radius
$R_3$ at $\sigma = \sigma_3$ and shrinks gradually as $\sigma$
increases, and stops contracting when $R = R_2$ at $\sigma = \sigma_2$.

Patching together two solutions in (\ref{ims}) we have again a
wormhole solution between a D3-brane and an anti-D3-brane. The distance
between the two three-branes is evaluated as $2\Delta \sigma$ with
\begin{equation}
\Delta \sigma = \sigma_2 - \sigma_3 = \la\sqrt{C}\left( I + 
\frac{3}{4\la^2Cf} J \right)
\label{del}\end{equation}
and
\begin{eqnarray}
I &=& \sqrt{R_2(R_2-R_1)}B\left(\frac{1}{2},\frac{1}{2}\right)F_1
\left(\frac{1}{2},-\frac{1}{2},-\frac{1}{2},1,-\frac{R_3-R_2}{R_2},
-\frac{R_3-R_2}{R_2-R_1}\right) \nonumber \\
&+& \frac{2R_1R_2}{\sqrt{R_3(R_2-R_1)}}
\Pi\left(\frac{\pi}{2},\frac{R_3-R_2}{R_3},k\right),  \\
J &=& \frac{2}{R_2\sqrt{R_3(R_2-R_1)}}\left(E - \frac{R_2}{R_1}
(E-K) \right),
\label{jek}\end{eqnarray}
where $k = \sqrt{(R_3-R_2)(-R_1)/(R_2-R_1)R_3}$ amd $F_1, K, E, \Pi$ are
Appell's hypergeometric function, the elliptic integral of the first,
second, third kind respectively. The expansion of $E$ in the first
term in (\ref{jek}) with respect to $k'=\sqrt{1-k^2}$ gives the
leading term $\Delta\sigma \simeq 3/2f$ in the small $f$ region,
which is the previous value in the linearized prescription.
Since the derivative of $R_3$ with respect to $z = f^2$ is negative for
$0 < \psi <\pi/2$
\begin{eqnarray}
\frac{dR_3}{dz} &=& - \left(\frac{3}{4\la^2C}\right)^{1/2}\frac{1}
{6z^{3/2}\sin \psi}(3\sin \psi(\cos\frac{\psi}{3} + 
\sqrt{3}\sin\frac{\psi}{3}) \nonumber \\
&+& 2\cos\psi (\sqrt{3}\cos\frac{\psi}{3} 
-\sin\frac{\psi}{3}) ),
\end{eqnarray}
the radius $R_3$ decreases under the increasing external field. On the 
other hand the radius $R_2$ increases because of 
\begin{equation}
\frac{dR_2}{dz} =  \left(\frac{3}{4\la^2C}\right)^{1/2}\frac{1}
{6z^{3/2}\sin \psi}F
\end{equation}
with 
\begin{equation}
F = 3\sin \psi(\sqrt{3} \sin\frac{\psi}{3} - \cos \frac{\psi}{3})
+ 2\cos\psi(\sin\frac{\psi}{3} + \sqrt{3}\cos\frac{\psi}{3}),
\end{equation}
whose positivity is shown as follows
\begin{eqnarray}
F &>& \sin\psi\left((3\sqrt{3}+2)\sin\frac{\psi}{3} + (2\sqrt{3}-3)
\cos\frac{\psi}{3}\right), \; \mathrm{for}\; 0 < \psi < \frac{\pi}{4},
\nonumber \\
F &>& \cos\psi \left((3\sqrt{3}+2)\sin\frac{\psi}{3} + (2\sqrt{3}-3)
\cos\frac{\psi}{3}\right), \; \mathrm{for}\; \frac{\pi}{4} \le \psi 
< \frac{\pi}{2}.
\end{eqnarray}
From these behaviors it follows that the radius $R_3$ coincides with
$R_2$, when the external field takes a critical value 
$f_c = (3/4\la^2C)^{1/4}$,  and then the funnel or wormhole
configuration is broken. 

Let us estimate the energy of funnel configuration extending from 
$\sigma_3$ to $\sigma_2$. Combining (\ref{ful}) with (\ref{cas})
we can extract it as 
\begin{equation}
E = T_1N\int_{\sigma_3}^{\sigma_2}d\sigma((1 + \la^2C(R')^2)
(\gamma fR^3 + 1) - \gamma R^3f)
\label{ext}\end{equation}
with $\gamma=4\la^2C/3$. Through the substitution of 
the minus expression in the first integral (\ref{fir}) into 
(\ref{ext}) the energy is given by
\begin{equation}
E = -T_1N \int_{\sigma_3}^{\sigma_2}d\sigma \left( 
\frac{4\la\sqrt{C}f}{3} R' \sqrt{-R^6 + \al R^4 
+ \beta R^3}((\frac{d\sigma}{dR})^2 + \la^2C ) \nonumber \\
 + \frac{4}{3}\la^2CR^3f \right).
\label{ene}\end{equation}
 The first term is rewritten as 
$T_1N\int_{\sigma_3}^{\sigma_2}d\sigma (\gamma fR^3 +1)$ whose first term
cancels the third term in (\ref{ene}). Thus we obtain 
in a suggestive form
\begin{equation}
E = -T_1N\la\sqrt{C}\gamma\int_{R_3}^{R_2} dR R^{3/2}
\sqrt{(R_3-R)(R-R_1)(R-R_2)} + T_1N\int_{\sigma_3}^{\sigma_2}d\sigma.
\label{sug}\end{equation}
The first term is further expressed as
\begin{eqnarray}
T_1N\la\sqrt{C}\gamma (R_3-R_2)^2R_2^{3/2}(R_2-R_1)^{1/2}
B\left(\frac{3}{2},\frac{3}{2}\right) \nonumber \\
\times F_1\left(\frac{3}{2},-\frac{3}{2},
-\frac{1}{2},3,-\frac{R_3-R_2}{R_2},-\frac{R_3-R_2}{R_2-R_1}\right)
\end{eqnarray}
and the second term is estimated by using (\ref{del}). Alternatively
we use a relation $T_1 = 2\pi\la T_3$ and the physical radius to
rewrite the first term in (\ref{sug}) as
\begin{equation}
T_3 \frac{N}{\sqrt{C}}\int_{\tilde{R}_2}^{\tilde{R}_3} d\tilde{R}
4\pi \tilde{R}^2\left( 1 - \frac{4f^2\tilde{R}^2}{9}
-\frac{2\la\sqrt{C}f}{3\tilde{R}}\right)^{1/2},
\end{equation}
which can be regarded as a bulk energy stored in the spherical 
D3-brane. The second term is naturally interpreted as the energy
of the $N$ D-strings with tension $T_1$ in the $X^9$ direction.
In view of the first integral (\ref{fir}) with $A=1$ and (\ref{rad})
we note that there is a simply expanded D-string solution with a
constant radius $R_2$ or $R_3$ for all $\sigma$ whose energy 
evaluated from (\ref{ful}) with $R'=0$ shows that it is a
nonmarginal bound state of $N$ D-strings and a D3-brane.
In the weak external field the total energy of the $R_2$ solution is
estimated to be lower than that of the $R_3$ solution.

\section{Conclusions}

In the framework of the nonabelian DBI theory we have observed that
there exists a nonocommutative stable solution describing 
the expanding of $N$ D0-branes into 
a  spherical D2-brane under a constant electric RR flux, even when
they sit in the curved near-horizon spacetime produced by the multiple
D4-branes. The same configuration appears also in the dual description 
using the abelian DBI action of a D2-brane with $N$ units of U(1) magnetic
flux. In Ref. \cite{RCM} the leading order terms in the nonabelian DBI 
action were analyzed for the multiple D0-branes in the flat spacetime,
while we have worked on the DBI action itself to see the dielectric 
effect in a curved spacetime. We have shown that the weak external 
field produces a stable solution for the near-horizon geometry.

In the behaviors of the noncommutative solution for the muliple D-strings
under the electric background flux we have seen that there are some
differences between the leading order prescription of the nonabelian
DBI action and the full one. The differences occur since the coupling
produced by the external flux breaks supersymmetries. 
In the absence of electric flux the
noncommutative BPS funnel solution has a pole where the multiple
D-strings expand into a spherical D3-brane with an infinite radius,
which is seen in both the leading order prescription and the full
DBI theory \cite{CMT}. When the non-zero external flux is turned on, this
pole structure is reproduced in the leading order prescription but
disappears in the full DBI theory. In no electric RR flux the 
expanded D-strings collapse down to zero size far away from the
position where the D3-brane opens up, while under non-zero flux they
shrink along the D-string worldsheet till having some minimum size.
The minimum sizes characterized by the strength of external field 
emerge in solving the full as well as linearized equations of motion
and take the same value in the weak external field limit.
Though the expanding of multiple D-strings occurs owing to both the
nonabelian nature of the world-volume theory and the effect of 
external electric flux, the minimum bound of puffing is 
produced by the latter effect. By solving the full equation of
motion we have seen that in the decreasing weak external field the minimum
size of spherical D3-brane decreases in proportion to the strength of
external field and the finite maximum size increases inversely in
proportion to it, whose behaviors are smoothly continued to those of
the BPS funnel solution in the zero external field limit.
By manipulating the full DBI action we can see the dependence of
the dielectric effect for the multiple D-strings on the strong
external field. As the external field becomes stronger, the slope of
our puffed funnel-like solution is reduced to be more gradual.
Specially at some critical strong external field the funnel
or wormhole configuration disappears. We guess that it may be 
related with the annihilation of a pair of D3-brane and
anti-D3-brane. 

The stable noncommutative expanded configuration of $N$ D0-branes
is regarded as a nonmarginal true bound state of $N$ D0-branes and a
spherical D2-brane, whose energy has a resemblance to that of the
nonmarginal BPS bound state of D$p$-branes and D$(p+2)$-branes 
compactified on a two-torus. The energy of the non-BPS funnel
configuration of $N$ D-strings has been shown to receive 
two separated contributions from the $N$ D-strings background and 
a spherical D3-brane. It is desirable to have a deeper understanding
why the similar separations emerge both in this non-BPS bound state
and in the marginal BPS bound state of Dp-branes and D$(p+4)$-branes
wrapped around a four-torus. 
One interesting extension of our work would be to investigate
the dielectric effect for the multiple D2-branes under the electric
RR flux and show how the expanding behaviors depend on the 
world-volume coordinates of D2-branes and  the strength of flux.

\end{document}